\documentclass{emulateapj}

\usepackage{amssymb}

\def\ts     {\thinspace}
\def\kms    {\ts km\ts s$^{-1}$}
\def\etal   {{\rm et\ts al.}}
\def\msol   {$M_{\odot}$}
\def\msolyr {$M_{\odot}$\,yr$^{-1}$}
\def\msolkpc {$M_{\odot}$\,kpc$^{-2}$}
\def\lsol   {$L_{\odot}$}
\def\lsolkpc {$L_{\odot}$\,kpc$^{-2}$}
\def\lprime {K\,\ts km\ts s$^{-1}$\,pc$^2$}

\def\aco    {{\rm CO}($J$=1$\to$0)}
\def\bco    {{\rm CO}($J$=2$\to$1)}
\def\bcoalt {{\rm CO} $J$=2$\to$1}

\def\eco    {{\rm CO}($J$=5$\to$4)}
\def\fco    {{\rm CO}($J$=6$\to$5)}

\slugcomment{draft version \today, accepted for publication in the Astrophysical Journal Letters}

\shorttitle{Molecular Gas in a $z$=5.3 Submillimeter Galaxy}
\shortauthors{Riechers et al.}

\begin{document}

\title{A Massive Molecular Gas Reservoir in the $z$=5.3 Submillimeter Galaxy AzTEC-3}

\author{Dominik A.\ Riechers\altaffilmark{1,7}, Peter L.\ Capak\altaffilmark{2},  Christopher L.\ Carilli\altaffilmark{3}, Pierre Cox\altaffilmark{4}, \\ Roberto Neri\altaffilmark{4}, Nicholas Z.\ Scoville\altaffilmark{1}, Eva Schinnerer\altaffilmark{5}, Frank Bertoldi\altaffilmark{6}, and Lin Yan\altaffilmark{2}}

\altaffiltext{1}{Astronomy Department, California Institute of
  Technology, MC 249-17, 1200 East California Boulevard, Pasadena, CA
  91125, USA; dr@caltech.edu}

\altaffiltext{2}{Spitzer Science Center, California Institute of Technology, MC 220-6, 1200 East California Boulevard, Pasadena, CA 91125, USA}

\altaffiltext{3}{National Radio Astronomy Observatory, PO Box O, Socorro, NM 87801, USA}

\altaffiltext{4}{Institut de RadioAstronomie Millim\'etrique, 300 Rue
  de la Piscine, Domaine Universitaire, 38406 Saint Martin d'H\'eres,
  France}

\altaffiltext{5}{Max-Planck-Institut f\"ur Astronomie, K\"onigstuhl 17, D-69117 Heidelberg, Germany}

\altaffiltext{6}{Argelander-Institut f\"ur Astronomie, Universit\"at
  Bonn, Auf dem H\"ugel 71, Bonn, D-53121, Germany}

\altaffiltext{7}{Hubble Fellow}


\begin{abstract}

We report the detection of \bcoalt, 5$\to$4, and 6$\to$5 emission in
the highest-redshift submillimeter galaxy (SMG) AzTEC-3 at $z$=5.298,
using the Expanded Very Large Array and the Plateau de Bure
Interferometer.  These observations ultimately confirm the redshift,
making AzTEC-3 the most submillimeter-luminous galaxy in a massive
$z$$\simeq$5.3 protocluster structure in the COSMOS field.  The
strength of the CO line emission reveals a large molecular gas
reservoir with a mass of 5.3$\times$10$^{10}\,(\alpha_{\rm
CO}/0.8)$\,\msol, which can maintain the intense 1800\,\msolyr\
starburst in this system for at least 30\,Myr, increasing the stellar
mass by up to a factor of six in the process.  This gas mass is
comparable to `typical' $z$$\sim$2 SMGs, and constitutes $\gtrsim$80\%
of the baryonic mass (gas+stars) and 30\%--80\% of the total
(dynamical) mass in this galaxy. The molecular gas reservoir has a
radius of $<$4\,kpc and likely consists of a `diffuse', low-excitation
component, containing (at least) 1/3 of the gas mass (depending on the
relative conversion factor $\alpha_{\rm CO}$), and a `dense',
high-excitation component, containing $\sim$2/3 of the mass.  The
likely presence of a substantial diffuse component besides
highly-excited gas suggests different properties between the
star-forming environments in $z$$>$4 SMGs and $z$$>$4 quasar host
galaxies, which perhaps trace different evolutionary stages.  The
discovery of a massive, metal-enriched gas reservoir in a SMG at the
heart of a large $z$=5.3 protocluster considerably enhances our
understanding of early massive galaxy formation, pushing back to a
cosmic epoch where the Universe was less than 1/12 of its present age.

\end{abstract}


\section{Introduction}

Our understanding of the physical properties of submillimeter galaxies
(SMGs; see review by Blain \etal\ \citeyear{bla02}) is of key
importance to studies of the early formation and evolution of massive
galaxies, as they are the likely progenitors of the most massive
galaxies in the present-day universe. SMGs typically represent
compact, intense ($>$1000\,\msol\,yr$^{-1}$), rather short-lived
($<$100\,Myr) starbursts with rapid gas consumption through high star
formation efficiencies that are commonly associated with ongoing major
mergers. Their star formation rates (SFRs) exceed those of `normal'
high-$z$ galaxies with comparable stellar mass ($M_\star$) by more
than an order of magnitude at $z$$\sim$2 (e.g., Daddi \etal\
\citeyear{dad09}), making them a comparatively rare, but
cosmologically important population. The extreme star formation events
in SMGs are typically associated with large amounts of gas and dust
which often obscure the stellar light and star formation at rest-frame
ultraviolet (UV) to optical wavelengths, making their identification
at such wavelengths notoriously difficult.

The most insightful way to study SMGs and their star formation
properties thus usually is through the dust-reprocessed UV light from
newly formed stars that is re-emitted in the far-infrared (FIR)
continuum (a measure of the SFR), and through line emission from
molecular gas (typically CO), the fuel for star fomation. Molecular
gas was detected in $>$20 SMGs to date, revealing large gas reservoirs
of $>$10$^{10}$\,\msol\ in most cases (see Solomon \& Vanden Bout
\citeyear{sv05} for a review).

Recently, Capak \etal\ (\citeyear{cap10}) discovered AzTEC-3, an SMG
at an unprecedented redshift of $z$=5.3. AzTEC-3 is not only the most
distant SMG known to date, but also resides in a massive, overdense,
proto-cluster environment extending out to $>$2\,Mpc, with two
companions within 12.2\,kpc. AzTEC-3 hosts a 1800\,\msol\,yr$^{-1}$
starburst\footnote{Assuming a Chabrier (\citeyear{cha03}) stellar
initial mass function (IMF).} exhibiting a FIR luminosity of $L_{\rm
FIR}$=(1.7$\pm$0.8)$\times$10$^{13}$\,\lsol\ that likely produced a
substantial fraction of its current stellar mass of
$M_\star$=(1.0$\pm$0.2)$\times$10$^{10}$\,\msol. The unusual nature of
AzTEC-3 and its potential major implications for massive galaxy
formation at very early cosmic epochs has initiated an in-depth study
of its physical properties and cosmic environment.

\begin{figure*}
\epsscale{1.15}
\plotone{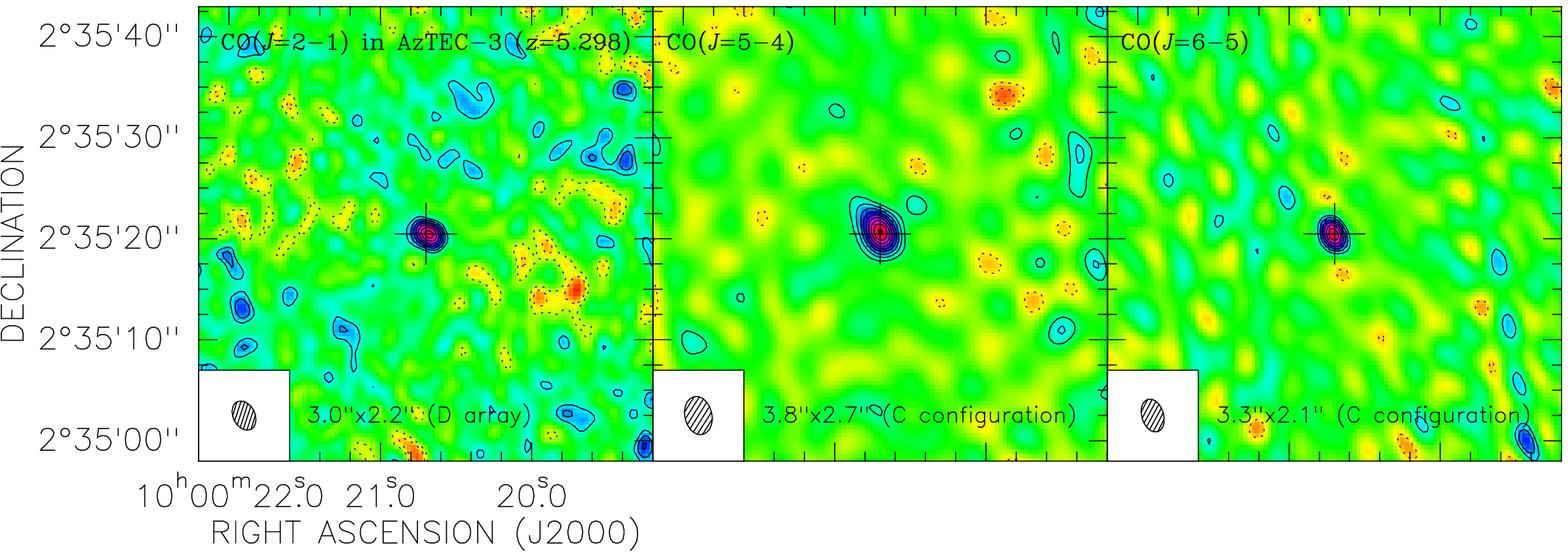}

\caption{Velocity-integrated EVLA/PdBI maps of the \bco\ ({\em left}), \eco\ ({\em middle}) and 
\fco\ ({\em right}) line emission over 950/852/874\,\kms\ toward AzTEC-3. At 
resolutions of 3.0$''$$\times$2.2$''$, 3.8$''$$\times$2.7$''$ and
3.3$''$$\times$2.1$''$ (as indicated in the bottom left corner of each
panel), the emission remains unresolved. The cross indicates the
position of the submillimeter continuum emission (Younger \etal\
\citeyear{you07}).  Contours are shown in steps of
1$\sigma$=0.045/0.10/0.22\,mJy\,beam$^{-1}$, starting at $\pm$2$\sigma$.
\label{f1}}
%
\end{figure*}

In this Letter, we report the detection of \bco, \eco\ and \fco\
emission toward AzTEC-3 ($z$=5.298), using the Expanded Very Large
Array (EVLA) and the Plateau de Bure Interferometer (PdBI).  We use a
concordance, flat $\Lambda$CDM cosmology throughout, with
$H_0$=71\,\kms\,Mpc$^{-1}$, $\Omega_{\rm M}$=0.27, and
$\Omega_{\Lambda}$=0.73 (Spergel \etal\ \citeyear{spe03},
\citeyear{spe07}).

\section{Observations}

\subsection{Plateau de Bure Interferometer}

We observed the \eco\ ($\nu_{\rm rest} = 576.2679305$\,GHz), and \fco\
(691.4730763\,GHz) emission lines toward AzTEC-3 using the PdBI.  At
$z$=5.298, these lines are redshifted to 91.5001 and 109.7925\,GHz
(3.3 and 2.7\,mm).  Observations were carried out under good 3\,mm
weather conditions in the 6C configuration on 2010 April 1 and 2,
resulting in 4.4 and 3.1\,hr on-source time for the CO $J$=5$\to$4 and
6$\to$5 lines, respectively. The nearby quasar B0906+015 (distance to
AzTEC-3:\ $12.8^\circ$) was observed every 22.5\,minutes for pointing,
secondary amplitude and phase calibration. For primary flux
calibration, the standard calibrators MWC349 and 3C84 were observed,
leading to a calibration that is accurate within $\lesssim$10\%.
Observations were set up using a total bandwidth of 1\,GHz (dual
polarization; corresponding to $\sim$3300/2700\,\kms\ at 3.3/2.7\,mm)
with the current correlator, and a total bandwidth of 3.6\,GHz (dual
polarization) with the new WideX correlator (recorded simultaneously).

For data reduction and analysis, the GILDAS package was used.  All
data were mapped using `natural' weighting.  The CO
$J$=5$\to$4/6$\to$5 data result in a final rms of
0.52/1.27\,mJy\,beam$^{-1}$ per 33/27\,\kms\ (10\,MHz) channel, and
0.05/0.13\,mJy\,beam$^{-1}$ (0.028/0.066\,mJy\,beam$^{-1}$) over the
entire 1\,GHz (3.6\,GHz) bandwidth. Maps of the velocity-integrated CO
$J$=5$\to$4/6$\to$5 line emission yield synthesized clean beam sizes
of 3.8$''$$\times$2.7$''$ and 3.3$''$$\times$2.1$''$ and rms noise
values of 0.10/0.22\,mJy\,beam$^{-1}$ over 852/874\,\kms\
(260/320\,MHz).

\subsection{Expanded Very Large Array}

We observed the \bco\ ($\nu_{\rm rest} = 230.53799$\,GHz) emission
line toward AzTEC-3 using the EVLA.  At $z$=5.298, this line is
redshifted to 36.6049\,GHz (8.2\,mm).  Observations were carried out
under good 9\,mm weather conditions in D array on 2010 May 24 and 30,
resulting in 6.3\,hr on-source time with 16\,antennas (equivalent to
2.2\,hr with 27\,antennas) after rejection of bad data. The nearby
quasar J1018+0530 (distance to AzTEC-3:\ $5.4^\circ$) was observed
every 7\,minutes for pointing, secondary amplitude and phase
calibration. For primary flux calibration, the standard calibrator
3C286 was observed, leading to a calibration that is accurate within
$\lesssim$10\%.  Observations were set up using a total bandwidth of
256\,MHz (dual polarization; corresponding to $\sim$2100\,\kms\ at
8.2\,mm) with the WIDAR correlator.

For data reduction and analysis, the AIPS package was used.  All data
were mapped using `natural' weighting.  The data result in a final rms
of 0.20\,mJy\,beam$^{-1}$ per 49\,\kms\ (6\,MHz) channel, and
0.030\,mJy\,beam$^{-1}$ over the entire 256\,MHz bandwidth. Maps of
the velocity-integrated CO $J$=2$\to$1 line emission yield a
synthesized clean beam size of 3.0$''$$\times$2.2$''$ at an rms noise
level of 0.045\,mJy\,beam$^{-1}$ over 950\,\kms\ (116\,MHz).

\begin{figure*}
\epsscale{1.0}
\plotone{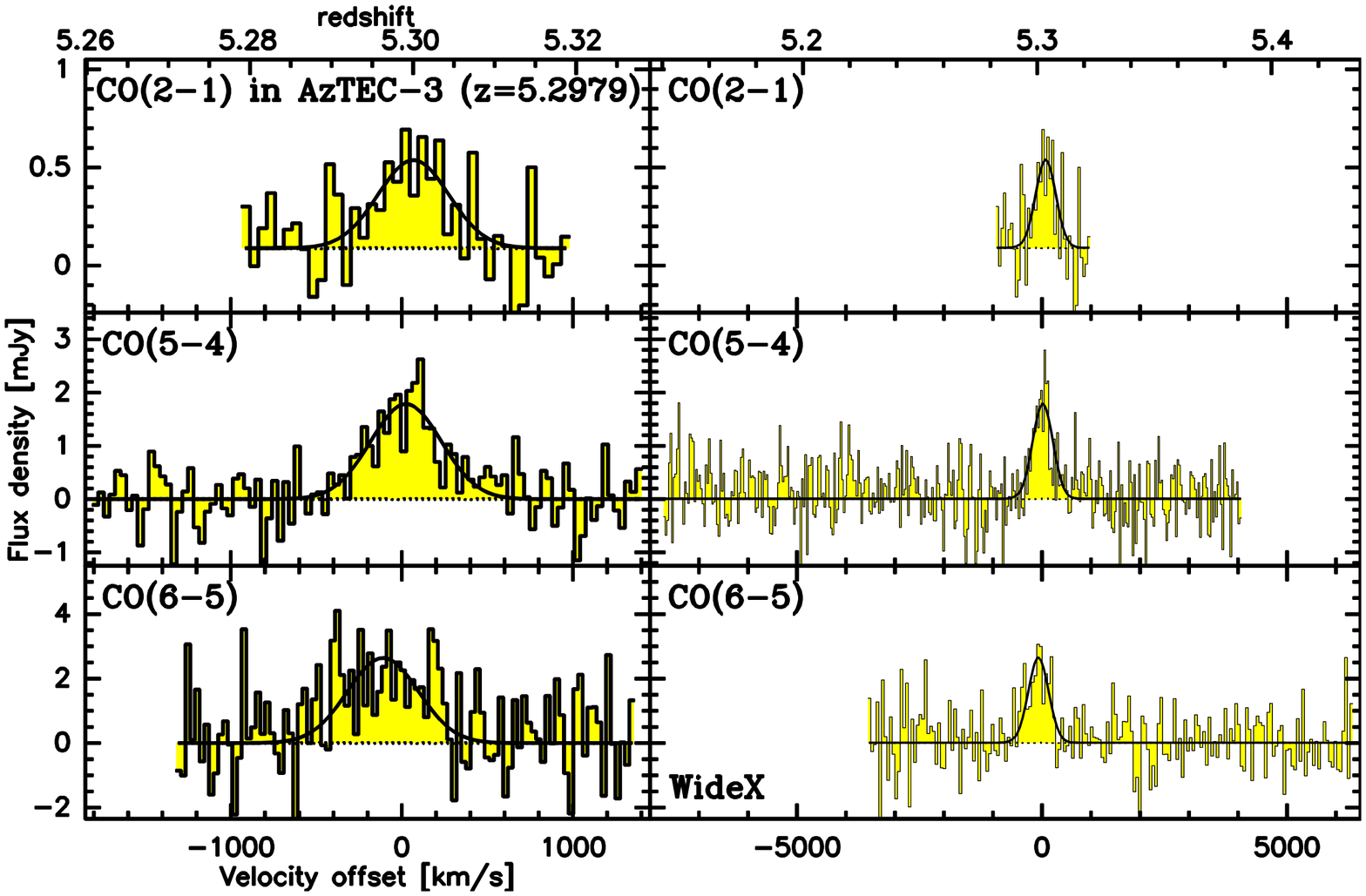}

\caption{{\em Left}:\ EVLA/PdBI \bco\ ({\em top}), \eco\ ({\em middle}) 
and \fco\ ({\em bottom}) spectra of AzTEC-3 at 6/10/10\,MHz
(49/33/27\,\kms ) resolution (histograms), along with Gaussian fits to
the line emission (black curves). The velocity scale is relative to
the source's redshift of $z$=5.2979$\pm$0.0004, as measured from the
molecular line emission.  {\em Right}:\ Same, but showing the PdBI
data recorded with the WideX correlator (CO $J$=6$\to$5 is re-binned
to 20\,MHz).
\label{f2}}
%
\end{figure*}

\section{Results}

We have detected \bco, \eco\ and \fco\ line emission toward the
$z$=5.298 SMG AzTEC-3 at 8$\sigma$, 10$\sigma$, and 7$\sigma$
significance (Fig.\ \ref{f1}). The CO $J$=5$\to$4 and 6$\to$5 data
yield a combined significance of 12$\sigma$.  We do not detect the
underlying continuum emission down to 2$\sigma$ limits of
0.13/0.31\,mJy\,beam$^{-1}$ (WideX correlator:\
0.08/0.14\,mJy\,beam$^{-1}$) at 3.3/2.7\,mm (rest-frame
520/434\,$\mu$m). A 2--3$\sigma$ peak is present in the 8.2\,mm map
(rest-frame 1.3\,mm), but offset from the CO peak position by
$\sim$1$''$. We thus consider the underlying continuum at this
wavelength undetected down to $\sim$0.09\,mJy\,beam$^{-1}$.

From Gaussian fitting to the line profiles (Fig.\ \ref{f2}), we obtain
CO $J$=2$\to$1, 5$\to$4 and 6$\to$5 line peak strengths of
0.45$\pm$0.07,\footnote{The fit also suggests an underlying continuum
component of 0.09$\pm$0.04\,mJy, consistent with the marginal peak
seen in the map.} 1.78$\pm$0.17 and 2.64$\pm$0.36\,mJy at a FWHM of
487$\pm$58\,\kms, centered at a (weighted median) redshift of
$z$=5.2979$\pm$0.0004 (consistent with the optical redshift; Capak
\etal\ \citeyear{cap10}). This corresponds to velocity-integrated emission
line strengths of 0.23$\pm$0.03, 0.92$\pm$0.09, and
1.36$\pm$0.19\,Jy\,\kms, i.e., line luminosities of $L'_{\rm
CO(2-1)}$=(5.84$\pm$0.78), $L'_{\rm CO(5-4)}$=(3.70$\pm$0.37), and
$L'_{\rm CO(6-5)}$=(3.82$\pm$0.54)$\times$10$^{10}$\,\lprime. This
also corresponds to CO $J$=6$\to$5/5$\to$4 and CO $J$=5$\to$4/2$\to$1
line brightness temperature ratios of $r_{65}$=1.03$\pm$0.16 (i.e.,
consistent with thermalized) and $r_{52}$=0.63$\pm$0.10.

\section{Analysis and Discussion}

\subsection{Origin of the CO and Submillimeter Emission}

\begin{figure}
\epsscale{1.15}
\plotone{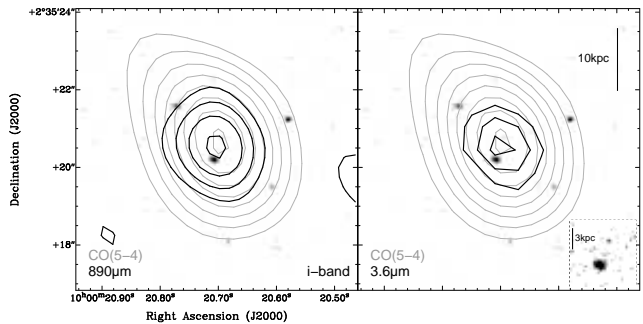}

\caption{Overlays of the \eco\ emission (gray contours) and
SMA 890\,$\mu$m ({\em left}; 2.7$''$$\times$2.1$''$ resolution;
Younger \etal\ \citeyear{you07}), as well as {\em Spitzer}/IRAC
3.6\,$\mu$m ({\em right}; PSF FWHM = 1.6$''$; from Sanders \etal\
\citeyear{san07}) emission (black contours in each panel), with an
{\em HST}/ACS $i$-band image (gray scale and inset; from Scoville
\etal\ \citeyear{sco07}).
\label{f3}}
%
\end{figure}

Within the size of the CO beam, there are three galaxies with colors
that are consistent with the CO redshift of AzTEC-3. By combining the
integrated emission of the CO $J$=5$\to$4 and 6$\to$5 lines detected
at 12$\sigma$ significance and fitting the $u-v$ data with a circular
Gaussian, we find that the source is unresolved down to a FWHM
diameter of 1.0$''$$\pm$0.7$''$ ($\sim$6$\pm$4\,kpc at $z$=5.3). This
is consistent with a limit of 1.3$''$$^{+0.9''}_{-1.3''}$
($\sim$8$^{+5}_{-8}$\,kpc) as derived from the CO $J$=2$\to$1 data.
Only one of the three color-selected galaxies falls within this
smaller area, identifying the $i$$\sim$26 galaxy targeted by the
optical spectroscopy as the real counterpart of the CO emission
(Fig.~\ref{f3}). The astrometric accuracy of the \bco\ and combined CO
$J$=5$\to$4 \& 6$\to$5 detections is 0.15$''$ and 0.12$''$. The peak
position of the CO emission is consistent with that of the 890\,$\mu$m
continuum emission (astrometric accuracy:\ 0.14$''$; Younger \etal\
\citeyear{you07}) and the 3.6\,$\mu$m emission (astrometric accuracy:\
0.2$''$; Sanders \etal\ \citeyear{san07}) within the relative
uncertainties (Fig.~\ref{f3}). Interestingly, the emission at all
these wavelengths is offset from the {\em HST}/ACS $i$-band image of
the source by $\sim$0.5$''$ (3\,kpc), but is consistent with a
marginal, much fainter peak in the rest-frame UV image (which we
interpret to be part of the same galaxy; inset in
Fig.~\ref{f3}). Given the 0.1$''$ relative astrometric accuracy of the
{\em HST} data, this offset is formally significant. We conclude that
this offset is due to a combination of lacking surface brightness
sensitivity and dust obscuration in the rest-frame UV image, a similar
effect as seen in the $z$=4.055 SMG GN20 (e.g., Carilli \etal\
\citeyear{car10}). This effect also explains the substantial
difference between the UV- and far-IR-derived star formation rates of
AzTEC-3 (Capak \etal\
\citeyear{cap10}), suggesting that the regions of most intense star
formation are highly dust-obscured.

\subsection{CO Line Excitation}

Based on the CO excitation ladder of AzTEC-3, we can constrain the
line radiative transfer through Large Velocity Gradient (LVG) models,
treating the gas kinetic temperature and density as free parameters.
For all calculations, the H$_2$ ortho--to--para ratio was fixed to
3:1, the cosmic microwave background temperature was fixed to 17.16\,K
(at $z$=5.298), and the Flower (\citeyear{flo01}) CO collision rates
were used. We adopted a CO abundance per velocity gradient of
[CO]/(${\rm d}v/{\rm d}r) = 1 \times 10^{-5}\,{\rm pc}\,$(\kms)$^{-1}$
(e.g., Wei\ss\ \etal\ \citeyear{wei05b},
\citeyear{wei07}; Riechers \etal\ \citeyear{rie06}). 

\begin{figure}
\epsscale{1.15}
\plotone{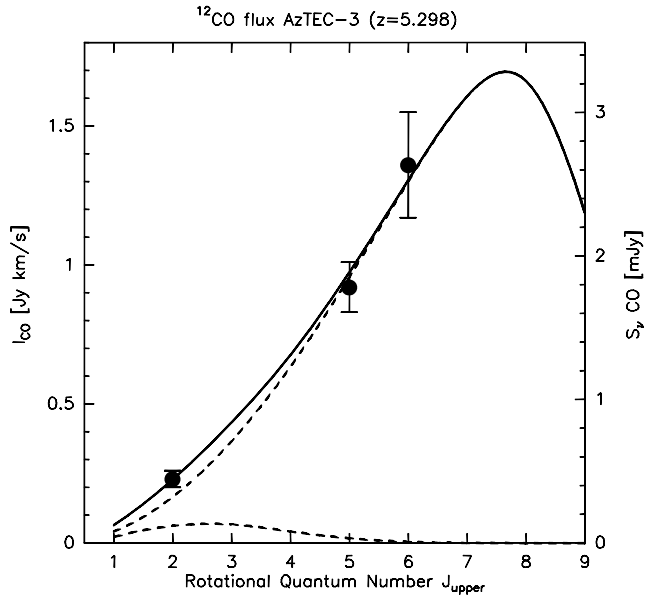}

\caption{CO excitation ladder (points) and LVG model (lines) for AzTEC-3. 
The model (solid line) consists of two gas components (dashed lines):\
a Milky-Way-like, low-excitation component with a kinetic temperature
of $T_{\rm kin}$=30\,K and a gas density of $\rho_{\rm
gas}$=10$^{2.5}$\,cm$^{-3}$, and a ULIRG-like, high-excitation
component with $T_{\rm kin}$=45\,K and $\rho_{\rm
gas}$=10$^{4.5}$\,cm$^{-3}$.
\label{f4}}
%
\end{figure}

The data are poorly fit by single-component models, which underpredict
the \bco\ flux by $\sim$30\%. The data can be fitted reasonably well
with two gas components, which are represented by a `diffuse',
low-excitation component with a kinetic temperature of $T_{\rm
kin}$=30\,K and a gas density of $\rho_{\rm
gas}$=10$^{2.5}$\,cm$^{-3}$, and a more `dense', high-excitation
component with $T_{\rm kin}$=45\,K and $\rho_{\rm
gas}$=10$^{4.5}$\,cm$^{-3}$ in our model (Fig.~\ref{f4}). Assuming
that the molecular gas is distributed in a face-on, circular disk, the
observed limits of 1.3$''$ and 1.0$''$ on the sizes of these low- and
high-excitation gas components yield CO disk filling factors of
$\gtrsim$75\% and $\gtrsim$10\%.  The `diffuse' gas component has
physical properties comparable to those of nearby spiral galaxies and
`normal' high-$z$ star-forming galaxies (e.g., Dannerbauer \etal\
\citeyear{dan09}), and contributes $\sim$35\% to the model-predicted
\aco\ flux. The `dense' gas component has properties comparable to
ultra-luminous infrared galaxy (ULIRG) nuclei and high-$z$
FIR-luminous quasars (e.g., Riechers et al.\ \citeyear{rie06},
\citeyear{rie09}), contributing $\sim$65\% to the
\aco\ flux. 

The model suggests an excitation-corrected \aco\ line luminosity of
$L'_{\rm CO}$=6.6$\times$10$^{10}$\,\lprime, and $L_{\rm
FIR}$/$L'_{\rm CO}$=260\,\lsol\,(\lprime )$^{-1}$. This luminosity
ratio is comparable to `typical' $z$$>$2 SMGs and quasar host galaxies
(e.g., Greve \etal\ \citeyear{gre05}; Riechers \etal\
\citeyear{rie06}).

\subsection{Total Molecular Gas Mass and Dynamical Mass}

Given the gas excitation conditions in AzTEC-3, we derive the total
molecular gas mass based on a ULIRG conversion factor of $\alpha_{\rm
CO}$=0.8\,\msol\,(\lprime )$^{-1}$ from $L'_{\rm CO}$ to $M_{\rm gas}$
(Downes \& Solomon \citeyear{ds98}), yielding $M_{\rm
gas}$=5.3$\times$10$^{10}$\,\msol.\footnote{Assuming a Milky-Way-like
$\alpha_{\rm CO}$=3.5\,\msol\,(\lprime )$^{-1}$ for the low-excitation
component would increase $M_{\rm gas}$ by a factor of 2.2.} This
corresponds to 5.3$\times$$M_\star$ in this system, implying that the
baryonic mass is dominated by the gaseous component. This means that
the intense, 1800\,\msol\,yr$^{-1}$ starburst in this galaxy has
sufficient fuel to more than sextuple $M_\star$ throughout the
remainder of its duration. This also sets the gas depletion timescale
$\tau_{\rm dep}$=$M_{\rm gas}$/SFR to 30\,Myr, comparable to lower
redshift SMGs (e.g., Greve \etal\
\citeyear{gre05}; Schinnerer \etal\ \citeyear{sch08}) and 
$z$$>$4 quasar host galaxies (e.g., Riechers \etal\
\citeyear{rie08}).

Adopting the size limit of 1.0$''$ ($\sim$6.2\,kpc), this also
corresponds to a limiting average gas surface density of $\Sigma_{\rm
gas}$$\gtrsim$1.7$\times$10$^9$\,\msolkpc. This is comparable to the
values found for $z$$\gtrsim$2 SMGs (typically
2$\times$10$^9$\,\msolkpc; Tacconi \etal\ \citeyear{tac06},
\citeyear{tac08}), and about half of the peak gas density of the
$z$=6.42 quasar J1148+5251 (3.5$\times$10$^9$\,\msolkpc; Riechers et
al.\ \citeyear{rie09}). Assuming an intrinsic $\Sigma_{\rm gas}$ in
this range (i.e., $\sim$2.7$\times$10$^9$\,\msolkpc ), this may
indicate that the gas reservoir in AzTEC-3 has an intrinsic circular
gas disk-equivalent radius of $r_0$$\sim$2.5\,kpc.

Based on the velocity and size information extracted from the CO data,
we can determine a value for the dynamical mass ($M_{\rm dyn}$) of
AzTEC-3. Given that the contributions from the central supermassive
black hole ($M_{\rm BH}$$\lesssim$0.1\% $M_\star$ $\propto$
10$^7$\,\msol; Alexander \etal\
\citeyear{ale08}) and dust ($M_{\rm dust}$$\lesssim$1--2\% $M_{\rm
gas}$; e.g., Michalowski et al.\ \citeyear{mic10}) to the total mass
budget are expected to be small, we here approximate $M_{\rm
tot}$$\simeq$$M_{\rm dyn}$$\simeq$$M_{\rm gas}$+$M_\star$+$M_{\rm
DM}$, where $M_{\rm DM}$ is the contribution from dark matter. We
further define the gas mass fraction as $f_{\rm gas}$=$M_{\rm
gas}$/$M_{\rm dyn}$ and the baryonic mass fraction as $f_{\rm
bary}$=1--$f_{\rm DM}$=[$M_{\rm gas}$+$M_\star$]/$M_{\rm dyn}$.

\begin{figure}
\epsscale{1.2}
\plotone{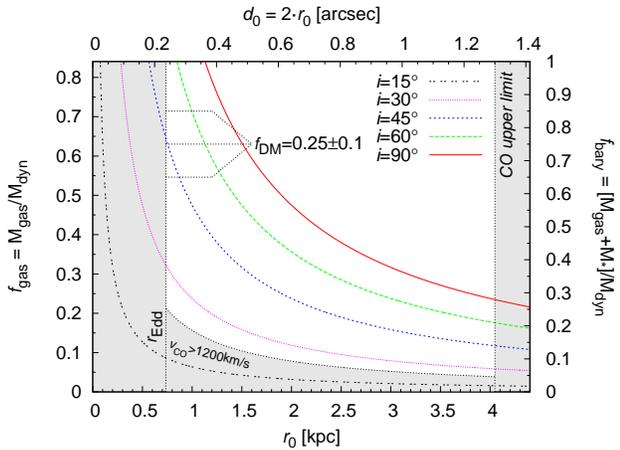}
\vspace{-7mm}

\caption{Constraints on the gas ($f_{\rm gas}$) and baryonic ($f_{\rm bary}$) 
 mass fractions for different line-of-sight inclination angles $i$ for
 AzTEC-3 (where $i$=90$^\circ$ corresponds to edge-on). $r_0$ is the
 equivalent radius of a circular disk used for deriving $M_{\rm
 dyn}$. The upper limit derived from the CO data and the lower limit
 derived from the Eddington limit on $L_{\rm FIR}$ ($r_{\rm Edd}$) are
 indicated by the vertical lines and shaded areas. A lower limit on
 $i$ is determined by assuming that $v_{\rm CO}$=d$v_{\rm
 FWHM}$\,sin$^{-1}$$i$$\leq$1200\,\kms, as indicated by the lower
 dotted curve and shaded area. The allowed range of $i$ for an assumed
 dark matter fraction of $f_{\rm DM}$=1$-$$f_{\rm bary}$=25\%$\pm$10\%
 (corresponding to $f_{\rm gas}$=63\%$\pm$8\%) is indicated by the
 thin dotted horizontal/arrow-shaped lines.
\label{f5}}
%
\end{figure}

In Fig.~\ref{f5}, $f_{\rm gas}$ and $f_{\rm bary}$ are shown as a
function of $r_0$ and the inclination $i$ of the gas disk. For
reference, $r_0$=2.5\,kpc corresponds to $M_{\rm
dyn}$\,sin$^2$$i$=1.4$\times$10$^{11}$\,\msol, i.e., $f_{\rm
gas}$$\sim$0.4 and $f_{\rm bary}$$\sim$0.45.\footnote{Assuming that
the system is seen close to edge-on.} From the observations of the gas
and dust in this galaxy, we can infer three main constraints on
$M_{\rm dyn}$ (shaded regions in Fig.~\ref{f5}). First, the CO
observations place an upper limit of 1.3$''$ on $r_0$. Second, of the
$>$20 SMGs detected in CO emission to date, none has a FWHM linewidth
of $>$1200\,\kms\ (median:\ 530\,\kms; Coppin \etal\
\citeyear{cop08}). Taking this as an upper limit on the intrinsic,
inclination-corrected linewidth $v_{\rm CO}$=d$v_{\rm
FWHM}$\,sin$^{-1}$$i$ yields a lower limit of $i$$>$24$^\circ$
($i$$=$0$^\circ$ corresponds to face-on). And third, assuming that the
starburst disk is supported by radiation pressure on dust grains
yields an Eddington limit on the FIR flux of $F_{\rm
FIR}$$\simeq$10$^{13}$\,\lsolkpc\ (e.g., Scoville
\citeyear{sco03}; Thompson \etal\ \citeyear{tho05}). Assuming that all 
$L_{\rm FIR}$ is due to star formation in the gas-rich reservoir, this
yields a lower limit on $r_0$ of $r_{\rm Edd}$=735\,pc. Note that the
corresponding dust brightness temperature limit of $T_{\rm dust}^{\rm
Edd}$=88\,K is more than twice as high as the $T_{\rm dust}$ obtained
from fitting the spectral energy distribution (Capak \etal\
\citeyear{cap10}), which may indicate that $F_{\rm FIR}$ in AzTEC-3 is 
well below the Eddington limit on average, contrary to what is seen,
e.g., in the center of the $z$=6.42 quasar J1148+5251 (Walter \etal\
\citeyear{wal09}; Riechers \etal\ \citeyear{rie09}).

Based on these constraints on $M_{\rm dyn}$, and assuming that
$\alpha_{\rm CO}$ and $M_\star$ are correct, another interesting limit
arises:\ for the galaxy not to be dark matter-dominated ($f_{\rm
bary}$$<$50\%) within its central few kpc, the system has to be seen
at $i$$>$35$^\circ$, and the starburst disk has to be relatively
compact ($r_0$$<$2.3\,kpc). Assuming a probably more realistic $f_{\rm
DM}$=25\%$\pm$10\% (e.g., Daddi et al.\ \citeyear{dad10}) would
suggest limiting $i$$\gtrsim$44$^\circ$$\pm$4$^\circ$ (at $r_{\rm
Edd}$) and $r_0$$\lesssim$1.5$\pm$0.2\,kpc (at $i$=90$^\circ$).

Overall, the observations thus may favor a relatively compact, highly
inclined galaxy, with a high, perhaps dominant, fraction of molecular
gas ($f_{\rm gas}$=0.3--0.8) that also dominates the baryonic mass in
this system ($f_{\rm gas}$$\simeq$0.84\,$f_{\rm bary}$). Despite the
limited spatial resolution of this detection experiment, the
diagnostic plot introduced here allows us to constrain the dynamical
mass in this system to within a factor of $\sim$2--3. However,
dynamically resolved CO observations at high ($<$0.7$''$) spatial
resolution are necessary to determine to what degree the assumptions
made here are correct. Also, potential effects of non-virial dynamics
(such as in a merger) require further investigation.

\section{Conclusions}

We have detected a molecular gas reservoir in the highest-redshift SMG
AzTEC-3 ($z$=5.298) that is comparable in mass to `typical' $z$$>$2
SMGs (e.g., Greve \etal\ \citeyear{gre05}; Ivison \etal\
\citeyear{ivi10}; Harris \etal\ \citeyear{har10}) and likely has a
low-excitation component similar\footnote{Future \aco\ observations
are desirable to characterize this component in more detail.}  in gas
properties to that seen in another $z$$>$4 SMG (Carilli \etal\
\citeyear{car10}), but in addition has a high-excitation gas component
with properties similar to $z$$>$4 quasar host galaxies (e.g.,
Riechers \etal\ \citeyear{rie06}, \citeyear{rie09}). The peak of the
gas distribution is coincident with the rest-frame optical and
far-infrared emission (stellar light and dust-enshrouded star
formation), but slightly offset from the peak of the rest-frame UV
emission (unobscured star formation), which appears to consist of
multiple bright clumps. Given the high gas mass fraction and star
formation rate of this galaxy, this suggests the presence of a heavily
obscured starburst, possibly triggered by a major, `wet' (i.e., very
gas-rich), $\lesssim$8\,kpc-scale merger.  This is reminiscent of what
is seen in `typical' $z$$>$2 SMGs (e.g., Tacconi \etal\
\citeyear{tac08}).

The detection of luminous CO emission implies relatively advanced
enrichment with heavy elements in the material that fuels the observed
early burst of star formation in AzTEC-3.  Assuming a Galactic
abundance of CO yields $M({\rm
CO})$$\simeq$7$\times$10$^7$\,\msol. Such a level of early metal
enrichment could be achieved through a few $\times$10$^7$ hundred
solar mass population-III stars (e.g., Walter \etal\
\citeyear{wal03}). This order-of-magnitude estimate, however,
corresponds to $\gtrsim$10\% $M_\star$, and would require a quite
`top-heavy' IMF.  We thus speculate that asymptotic giant branch stars
and supernovae may be extremely efficient in enriching their
environments at such early epochs, or that the stellar mass may be
underpredicted due to obscuration in the most intensely star-forming
regions.

The unusually high-excitation gas component in AzTEC-3 (for a SMG)
raises the question whether or not the environment may play a role in
determining its star formation properties. AzTEC-3 has two close,
massive companions with consistent photometric redshifts within only
12.2\,kpc. The molecular gas reservoir is clearly spatially separated
from the companions; however, gravitational interactions may still
play a role at such small distances, which may explain an above
average peak gas density (leading to high CO excitation).  The most
similar object probably is the $z$=4.055 SMG GN20, which evolves in a
comparably overdense environment (Daddi \etal\
\citeyear{dad09}). Interestingly, the CO reservoir in AzTEC-3 is more
compact, (partially) higher excited, and less than half as massive as
that in GN20. This may either indicate that the (overdense) cosmic
environments of both galaxies are considerably different (GN20 even
has two massive SMG companions within 180\,kpc distance), or that they
are in different stages of their evolution as a SMG.  The likely
presence of a substantial low-excitation gas component besides
highly-excited gas in both systems may point at a fundamental
difference in physical properties between the star-forming
environments in $z$$>$4 SMGs and $z$$>$4 FIR-luminous quasar host
galaxies (e.g., Riechers \etal\ \citeyear{rie06}), which may trace
overall different evolutionary stages of galaxies with comparable,
high gas masses and star formation rates. This difference may arise
due to a higher average concentration of the gas in the quasar hosts
(rather than large-scale influence of the active nucleus), e.g., due
to later merger stages. This would be consistent with a scenario where
SMGs evolve into FIR-luminous quasars (e.g., Sanders \etal\
\citeyear{san88}; Coppin \etal\ \citeyear{cop08}). Both populations 
are likely progenitors of the most massive galaxies in the present-day
universe.

This investigation clearly motivates more detailed studies of the
first SMGs and quasars, which will enable us to directly probe the
scales that are critical to unravel the physical processes that drive
the clustered evolution of massive galaxies in the early universe,
back to the first billion years after the Big Bang.

\acknowledgments 
We thank Josh Younger for the SMA image of AzTEC-3, Christian Henkel
for the original version of the LVG code, and the referee for a
helpful report. DR acknowledges support from from NASA through Hubble
Fellowship grant HST-HF-51235.01 awarded by STScI, operated by AURA
for NASA, under contract NAS\,5-26555.  The IRAM PdBI is supported by
INSU/CNRS (France), MPG (Germany) and IGN (Spain). The EVLA is a
facility of NRAO, operated by AUI, under a cooperative agreement with
the NSF.

\end{document}